\documentclass[11pt]{article} 
\usepackage{fancybox}
\usepackage{graphicx}
\usepackage{epsf}
\usepackage{amsmath,amssymb}
\usepackage{url}
\usepackage{hyperref}
\usepackage{color}
\definecolor{rosso}{cmyk}{0,1,1,0.4}
\definecolor{rossos}{cmyk}{0,1,1,0.55}
\definecolor{rossoc}{cmyk}{0,1,1,0.2}
\definecolor{blus}{cmyk}{1,1,0,0.6}
\definecolor{blum}{cmyk}{1,1,0,0.5}
\definecolor{blu}{cmyk}{1,1,0,0.4}
\definecolor{bluc}{cmyk}{1,1,0,0.1}
\definecolor{verde}{cmyk}{0.92,0,0.59,0.25}
\definecolor{verdec}{cmyk}{0.92,0,0.59,0.15}
\definecolor{verdes}{cmyk}{0.92,0,0.59,0.4}
\definecolor{giallo}{cmyk}{0,0,1,0}
\definecolor{gialloverde}{cmyk}{0.44,0,0.74,0}
\usepackage{setspace}
\begin{document} 
\thispagestyle{empty}
\vspace{20mm}
\begin{center}
{\LARGE{\bf The Higgs boson and}}\\[3mm] 
{\LARGE{\bf the International Linear Collider}}
\\[20mm]
{\large {\bf Francesca~Borzumati$^{1,2,3}$ and Eriko Kato$^{2}$}}
\end{center}
\begin{center}
{\emph{
$^1$Institute for International Education, 
 Tohoku University,  Aoba ku, Sendai, Japan \\[1.5ex]
$^2$Physics Department, Tohoku University, 
 Aramaki, Sendai, Japan  \\[1.5ex]
$^3$ Scuola Internazionale di Studi Superiori Avanzati,
Trieste, Italy
}}
\end{center}\vskip 2cm

\setstretch{1.08}
\begin{abstract}
The Higgs boson will be subject of intense experimental searches
 in future high-energy experiments. 
In addition to the effort made at the Large Hadron Collider, where 
 it was discovered, it will be the major subject of study at the
 International Linear Collider. 
We review here the reasons for that and some of the issues to be 
 tackled at this future accelerator, in particular that of the 
 precision of the Higgs-boson couplings. 
\end{abstract}
\vfill 
\eject

The Large Hadron Collider~(LHC) at CERN has brought physics to the 
 TeV energy frontier. 
It has been a long-standing belief that at this energy an important 
 issue regarding the Standard Model~(SM) of particle physics would be
 answered and New Physics~(NP) (as in addition to that described by 
 the SM) would be observed. 

In its first stage, at 7 and 8~TeV of center of mass~(CM) energy, 
 however, no NP has been found, leaving physicists 
 uncomfortably bewildered about their long-held view that TeV- or  
 even subTeV-NP is responsible for the stabilization of the Fermi
 scale and of the Higgs mass. 
While the final answer to this issue has still to be given by the
 second stage of the LHC at CM energy of 14~TeV, at least one part 
 of physicists expectations was realized: 
 in July 2012, the discovery of a new boson was 
 announced~\cite{EXPdiscov1,EXPdiscov2}. 
In the following months, it has became increasingly clear that this 
 particle is indeed the Higgs boson responsible for the breaking of 
 the SM gauge symmetry and for providing mass to the SM elementary 
 particles. 

The frantic activity to fingerprint this particle will, without
 any doubt, increase in the second LHC stage. 
In case NP turns out to be more elusive than expected, the 
 measurement of the Higgs-boson couplings can already give us information
 on what we are to find. 
Any top-like NP state $X$ capable of stabilizing the Fermi scale and the 
 Higgs mass itself, for example, is bound to produce a deviation of the
 couplings Higgs-gluon-gluon $g_{Hgg}$ and 
 Higgs-photon-photon $g_{H\gamma\gamma}$, which can be numerically
 expressed as\footnote{All the following estimates are
 from Baer et al.~\cite{TDRphys}, where the value of 120~GeV
 was chosen for the Higgs-boson mass. An update of all previous
 studies, using $m_H = 126\,$GeV is still missing.}:
\begin{equation}
\frac{g_{Hgg}}{g_{Hgg}\vert_{SM}}  = 
 1 + 1.4\% \left(\frac{\rm TeV}{m_X}\right)^2,
\hspace*{2.5truecm}
\frac{g_{H\gamma\gamma}}{g_{H\gamma\gamma}\vert_{SM}}  = 
  1 -0.4\%\left(\frac{\rm TeV}{m_X}\right)^2,
\label{scalDev}
\end{equation}
if the state $X$ is a top-like scalar, or 
\begin{equation}
\frac{g_{Hgg}}{g_{Hgg}\vert_{SM}}  =  
  1 + 2.9\%\left(\frac{\rm TeV}{m_X}\right)^2,
\hspace*{2.5truecm}
\frac{g_{H\gamma\gamma}}{g_{H\gamma\gamma}\vert_{SM}}  =  
  1 -0.8\%\left(\frac{\rm TeV}{m_X}\right)^2,
\label{fermDev}
\end{equation}
if it is a top-like fermion. 
Thus, a resolution of the couplings $g_{Hgg}$ and
 $g_{H\gamma\gamma}$ at the percent level is needed to tell us
 something about this possible new state $X$.

Supersymmetric particles, if Supersymmetry~(SUSY) is the 
 theory that extends the SM, produce numerically similar 
 deviations of the $g_{Hgg}$ and $g_{H\gamma\gamma}$ couplings,
 in the minimal SUSY SM with $m_A = 1\,$TeV, $\tan \beta = 1$, 
 and stop particles with masses $860$ and $1200\,$GeV:
\begin{equation}
\frac{g_{Hgg}}{g_{Hgg}\vert_{SM}}  =  
  1 - 2.7, 
\hspace*{3.5truecm}
\frac{g_{H\gamma\gamma}}{g_{H\gamma\gamma}\vert_{SM}}  =  
  1 +0.2,
\label{SUSYdev}
\end{equation}
 with a peculiar change of signs with respect to those 
 obtained in Equations~(\ref{scalDev}),~(\ref{fermDev}).
This is due to the fact that in the SUSY situation
 charginos/charged Higgs/staus contribute only to the  
 $g_{H\gamma\gamma}$ coupling, whereas stops contribute to both. 

Larger deviations can be obtained in other models, 
 but it is clear that to disentangle at least the simple
 extensions discussed above, a precision at the percent level, 
 or even smaller is required.
It is then clear why the measurement 
 precision of the Higgs couplings has become an arena for 
 debate on which new accelerators, if any, should 
 guide us into future explorations.

The importance of $e^+e^-$ machines for these future 
 explorations of particle physics has long been discussed
 (see for example~\cite{EKato},~\cite{LCPREPORT}.)

Lepton $e^+e^-$ colliders have quite some advantage over 
 hadron colliders when it comes to precision measurements. 

In a hadron machine, as the LHC, about 30 $pp$ collisions
 take place at each bunch crossing with each of them producing
 hundreds of particles. 
At an $e^+e^-$ collider only one photon-photon collision is
 expected at each bunch crossing. 
Thus, detectors for $e^+e^-$ machines are unburdened from 
 high-occupancy problems. 
They can be as thin as possible and 
 physically located much closer to the interaction point. 
In turn, this allows a factor of 10 improvement in momentum 
 resolution, with respect to a hadron collider like the LHC, 
 a factor of about 3 improvement in jet-energy resolution, 
 and, last, but not least, also an excellent capability of 
 tagging $b$- and $c$-quarks, as well as $\tau$-leptons. 

Together with the fact that pileups from multiple collisions
 are not a problem in an $e^+e^-$ environment, this makes clear
 why an $e^+e^-$ machine is more suited for precision 
 measurements. 

The unique feature of the International Linear Collider~(ILC)
 is the fact that its CM energy can be increased gradually 
 simply by extending the main linac. 

The ILC is supposed to operate at (at least) three stages. 
After a start at $\sqrt{s}=250\,$GeV, there will be an increase
 of the CM-energy to $500\,$GeV, and later to $1000\,$GeV. 
In the three stages, each one operating for three years, an 
 integrated luminosity of $250$, $500$, and $1000\,$fb$^{-1}$ 
 will be obtained, respectively. 
With a luminosity upgrade, the accumulated luminosity will 
 reach values of $1150$, $1600$, and $2500\,$fb$^{-1}$, 
 respectively. 
Given these luminosities, it is  
 reasonable to assume that at the $250\,$ and $500\,$GeV ILC, 
 about 80 thousand and 120 thousand Higgs events will be produced.
With the luminosity achieved after a luminosity upgrade, this 
 amount will be at least tripled.

\begin{figure}[hb]
\begin{center}
\includegraphics[width=0.30\textwidth]{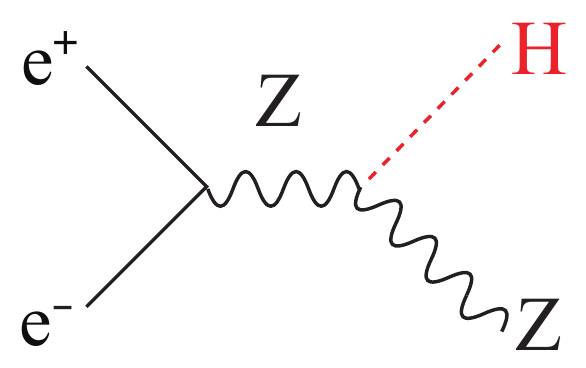}
\hspace*{0.5truecm}
\includegraphics[width=0.30\textwidth]{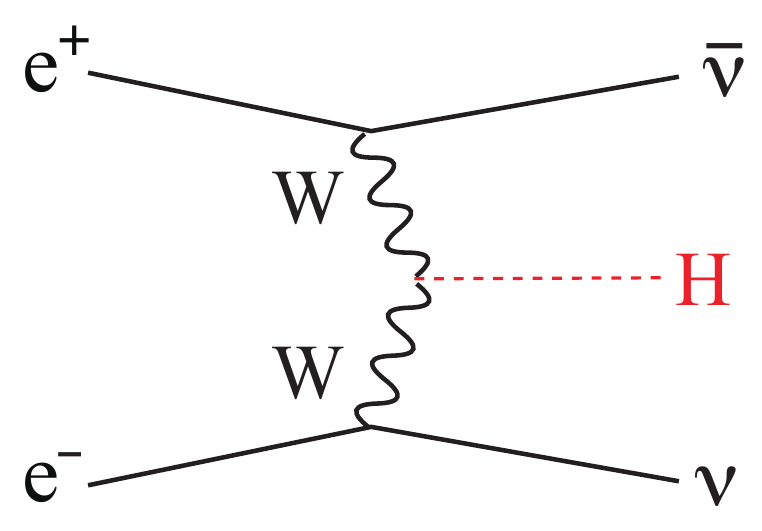}
\end{center}
\caption{\it Feynman diagrams for the dominant Higgs-production processes
 at the ILC: the Higgs-strahlung~(left) and the $WW$-fusion
 processes~(right).}
\label{fig:FeynDZHWH}
\end{figure}
At the different stages the Higgs couplings are measured in different 
 processes. 
The two main Higgs-production mechanisms at the ILC are
 Higgs-strahlung, {\it i.e.} the radiation of the Higgs bosons off an
 $s$-channel $Z$-line, and the $WW$-fusion process. 
The two mechanisms are depicted diagrammatically in 
 {\bf Figure~\ref{fig:FeynDZHWH}}.
A replacement of the $W$ boson with the $Z$ boson and of the final
 neutrino and antineutrino with an electron and a positron in the
 diagram for the $WW$-fusion mechanism  is the diagram for the
 sub-leading $ZZ$-fusion production mechanism. 

\begin{figure}[t]
\begin{center}
\includegraphics[width=0.55\textwidth]{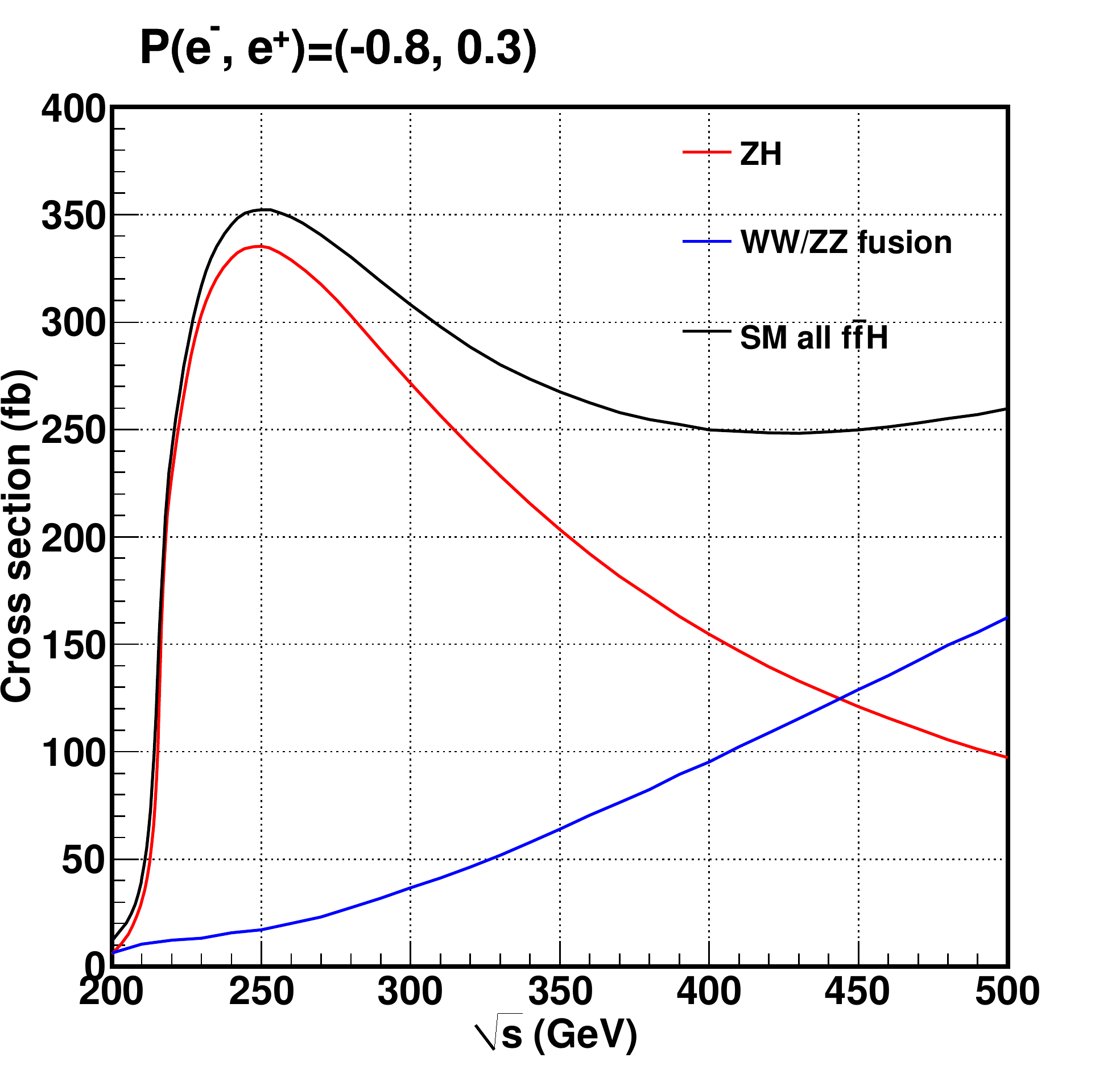}
\end{center}
\caption{\it Cross sections for the Higgs-strahlung (red line) 
 and $WW$-fusion (blue line) production mechanism, as a function of the CM 
 energy~$\sqrt{s}$ ~\cite{TDRphys}.
 The black line is simply the cross section due to both
 production mechanisms, giving rise to the final state $H f \bar{f}$
 (with the $Z$-boson in the Higgs-strahlung cross section decaying in all 
 possible fermions). 
 Polarizations of $80\%$ and $30\%$ for electrons and positrons,
 respectively, were assumed.}
\label{fig:XsecZHWH}
\end{figure}
As shown by the values of cross sections for the first two 
 production mechanisms, in {\bf Figure~\ref{fig:XsecZHWH}},
 the Higgs-strahlung process peaks at $\sqrt{s}=250\,$GeV,
 while at higher energies, the dominant
 production process is the $WW$-fusion process.
The blue line in this figure includes
 also the cross section for the $ZZ$-fusion process, which is only
 a tiny fraction of that for the $WW$-fusion process. 

With the subsequent decays of the Higgs boson 
 into a pair of $X$ particles, $H\to XX$, these two processes 
 allow to study the couplings $g_{HXX}$, which can be tree-level 
 couplings, or also effective couplings for loop-level decays. 
Indeed, the particle $X$ here can be a fermion,
 a photon or a gluon.

The $ZH$ process is crucial for measuring the Higgs couplings. 
By reconstructing the $Z$ boson from the two leptons to which it 
 decays, the Higgs boson can be captured inclusively, without 
 looking at its decay products.
Thus, also the width for invisible decays can be estimated. 

With the Higgs recoil method, and the luminosity values specified above,
 the total cross section for the 
 $ZH$ process, $\sigma_{ZH}$,  can be measured with $1.3\%$ precision.

Measuring then the cross section for the process 
 $e^+e^- \to ZH \to ZXX$, $\sigma_{ZH;H \to XX}$, sensitive to 
 the coupling $g_{HXX}$:
\begin{equation}
 \frac{\sigma_{ZH;H \to XX}}{\sigma_{ZH}} \propto g_{HXX}^2
\label{eq:gHXX}
\end{equation}
 a good precision can be achieved for the couplings $g_{HXX}$.
In particular a precision of $0.7\%$ can be obtained in the 
 case $X=Z$, i.e. for  $g_{HZZ}$, when $m_H=126\,$GeV is used. 
(Note the claimed precision of $0.5\%$ listed in
 {\bf Table~\ref{tab:coup_indep}}, referring to $m_H=120\,$GeV.)

The most challenging decays among those studied at this CM energy, 
 are those into $b$-, $c$- quarks, and into gluons.

\begin{table}[t]
\begin{center}
\begin{tabular}{lcccc}
 & ILC(250)     & ILC(500) & ILC(1000)  & ILC(LumUp) \cr 
$\sqrt{s}$~(GeV)&    250   &   250+500  & 250+500+1000 & 250+500+1000  \cr 
  L  (ab$^{-1}$) &   0.25   &  0.25+0.5  & 0.25+0.5+1   & 1.15+1.6+2.5 \cr  
\cline{1-5}
$\gamma\gamma$  & 18  \%    & 8.4  \%    & 4.0  \%      & 2.4 \%  \cr
$gg$            & 6.4 \%   & 2.3  \%    & 1.6  \%      & 0.9 \%  \cr
$WW$            & 4.8 \%   & 1.1  \%    & 1.1  \%      & 0.6 \%  \cr
$ZZ$            & 1.3 \%   & 1.0  \%    & 1.0  \%      & 0.5 \%  \cr
$t\bar t$       & --       & 14   \%    & 3.1  \%      & 1.9 \%  \cr
$b\bar b$       & 5.3 \%   & 1.6  \%    & 1.3  \%      & 0.7 \%  \cr
$\tau^+\tau^-$  & 5.7 \%   & 2.3  \%     & 1.6  \%      & 0.9 \%  \cr
$c\bar c$       & 6.8 \%   & 2.8  \%    & 1.8  \%      & 1.0 \%  \cr
$\mu^+\mu^-$    & 91\%     & 91\%        & 16 \%        & 10 \%  \cr
$\Gamma_T(h)$   & 12 \%    & 4.9 \%      & 4.5 \%       & 2.3  \%  \cr
$HHH$           & --       & 83 \%       & 21 \%        & 13 \%  \cr 
BR(invis.)      &$<$ 0.9\% &$<$ 0.9\%    &$<$ 0.9\%    &$<$ 0.4\%  \cr
\end{tabular}
\caption{Expected precisions for the Higgs-boson
  couplings~\cite{Asner2013psa}, once results from the different 
  CM energies, and upgraded luminosity are considered.
 For these estimates the value $m_H = 120\,$GeV was used.}
\label{tab:coup_indep}
\end{center}
\end{table} 

The expected sensitivity for the couplings measured at this stage is
 summarized in {\bf Table~\ref{tab:coup_indep}}, where also the couplings
 $g_{HWW}$, $g_{Htt}$, measured at $500\,$GeV, and discussed hereafter,
 are included.

At $\sqrt{s}=500\,$GeV, the measurement of the cross section
 for the process
 $e^+e^- \to \nu \bar{\nu}H \to \nu \bar{\nu}WW$,
 $\sigma_{\nu \bar{\nu} H;H \to W W}$, related to 
 $g_{HWW}$ as in 
\begin{equation}
 \sigma_{\nu \bar{\nu}H;H \to WW} =
    \sigma_{\nu\bar{\nu} H} BR(H\to WW) \propto g_{HWW}^4
\label{eq:gHWW}
\end{equation}
 gives this coupling, up to the absolute cross section $\sigma_{\nu\nu H}$.
This can be obtained from the measurement of 
 $\sigma_{\nu \bar{\nu}H;H \to XX}$:
\begin{equation}
 \sigma_{\nu \bar{\nu}H;H \to XX} =
    \sigma_{\nu\bar{\nu} H} BR(H\to XX) \propto g_{HWW}^2 g_{HXX}^2 
 \label{eq:vvHXX}
\end{equation}
 with the coupling $g_{HXX}^2$ already extracted at $\sqrt{s}=250\,$GeV.
A convenient decay mode is in this case $H\to b\bar{b}$.
\begin{figure}[h]
\begin{center}
\includegraphics[width=0.26\textwidth]{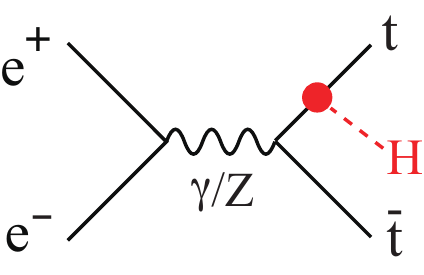}
\hspace*{0.5truecm}
\includegraphics[width=0.26\textwidth]{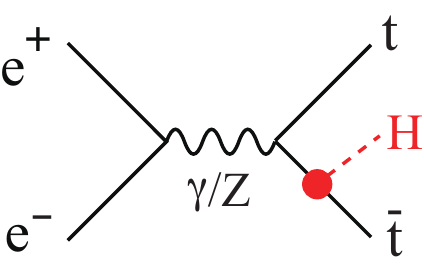}
\hspace*{0.5truecm}
\includegraphics[width=0.26\textwidth]{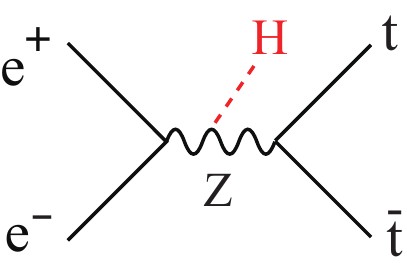}
\end{center}
\caption{\it Feynman diagrams for the $t\bar{t}H$ production process,
 with the Higgs boson radiated off the $t$-quark and the $Z$-line.}
\label{fig:FeynDttH}
\end{figure}

At $\sqrt{s}=500\,$GeV also the $t\bar{t}H$-production
 mechanism opens up.  
The relevant Feynman diagrams for this process are shown in 
 {\bf Figure~\ref{fig:FeynDttH}}.

\begin{figure}[h]
\begin{center}
\includegraphics[width=0.9\textwidth]{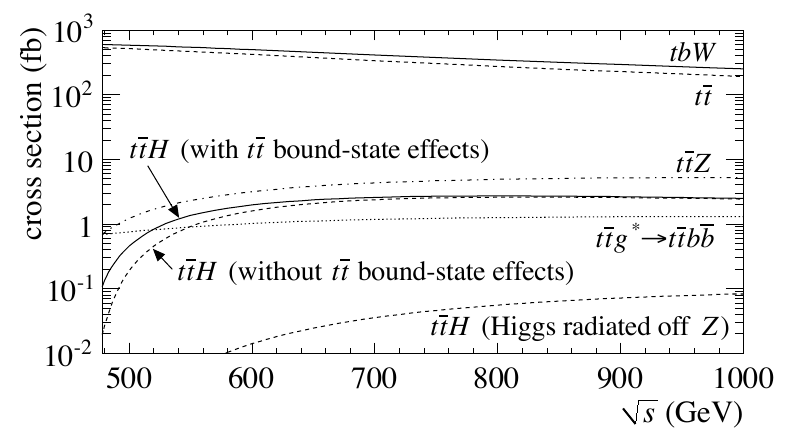}
\end{center}
\caption{\it Cross sections for the $t\bar{t}H$
 process as a function of $\sqrt{s}$~\cite{Yonamine2011jg}.
The solid~(dotted) lines highlighted by arrows, denote which
 one  includes (does not include) QCD effects. Both lines refer to the 
 process shown in {\bf Figure~\ref{fig:FeynDttH}}.
 For completeness, also the cross section for the $t\bar{t}H$ process 
 originating from teh HIggs strahlung, with subsequent decay $H\to t \bar{t}$
 is given.
 Shown are also some sources of background, such as the process 
 $t\bar{t}g^\ast \to t \bar{t} b\bar{b}$ with a virtual gluon radiated off 
 one of the two initial top-quarks.}
\label{fig:XsecttH}
\end{figure}

The corresponding cross section 
is plotted in {\bf Figure~\ref{fig:XsecttH}} as a function of the 
 CM energy. 
At $\sqrt{s}=500\,$GeV the cross section is somewhat small. It is
 however enhanced by a factor of two by QCD
 corrections~\cite{Yonamine2011jg}, including 
 mainly $t\bar{t}$ bound-state effects. This makes the 
 measurement of the Higgs-top coupling possible, with a somewhat 
 modest precision of $14\%$, not including theoretical errors.
Subsequent CM-energy increases and luminosity upgrades will 
 be able to improve this precision up to $2\%$ (again without
 theory errors).

Data at $500$ and $1000\,$GeV will help improving the precision
of all couplings already measured in previous stages. 

What will probably remain a bit elusive, probably until the 
 completion of the ILC program, is the Higgs self coupling,
 $g_{HHH}$. 
It is clear that to give a complete identification of the 
 recently discovered Higgs boson the measurement of this 
 self coupling and then a reconstruction of the Higgs potential
 are necessary. 

It can be obtained from the $ZH$ and $WW$-fusion processes, with 
 a subsequent decay of $H$ into a pair $HH$.
In the $e^+e^- \to ZHH$ process the $Z$ boson will be reconstructed 
 from the pairs $l \bar{l}$ and $q \bar{q}$ in which it decays. 
In the $e^+e^- \to \nu \bar{\nu} HH$ the two $H$'s can be 
 reconstructed from $4b$'s or $2b$'s and 2 $W$'s. 

Luminosity upgrade at the ILC as well as high polarization will 
 have the goal of bringing the measurement of the Higgs self coupling 
 to a final $\sim 10\%$ precision.

Except for this coupling, the precision achievable for the other 
 couplings, summarized in {\bf Table~\ref{tab:coup_indep}} is well 
 within the values requested by
 Equations~(\ref{scalDev},~\ref{fermDev},~\ref{SUSYdev}), {\it i.e.} 
 is good enough to address the issue of the nature of NP to be
 expected.

The LHC will in the meantime help (hopefully even considerably) 
 in fingerprinting the Higgs boson. 
Together with the ILC, then, an accurate profile of this particle
 will be put together and, with it, an understanding of what triggers 
 the electroweak symmetry breaking will be obtained. 

\vspace*{1truecm}
\large {\bf Acknowledgements}
The work of Francesca~Borzumati is partially supported by the
 Grant-in-Aid for Scientific Research~23540283 of JSPS, Japan, and
 by the ERC Advanced Grant 267985 ``DaMESyFla''.
Eriko~Kato is supported by the JSPS Grant in Aid for Specially-Promoted
Research ``A global R$\&$D program of a state-of-the-art detector
system for ILC''.

\end{document}